# *A Novel Formal Agent-based Simulation Modeling Framework of an AIDS Complex Adaptive System*

Amnah Siddiqa[1], Muaz Niazi[2,]*

[1]*Research Centre for Modeling and Simulation, National University of Sciences and Technology, Islamabad, Pakistan*

[2]*Office of Research, Innovation & Commercialization, Bahria University, Islamabad, Pakistan*

*Corresponding author muaz.niazi@gmail.com

## *Abstract*

HIV/AIDS spread depends upon complex patterns of interaction among various sub-sets emerging at population level. This added complexity makes it difficult to study and model AIDS and its dynamics. AIDS is therefore a natural candidate to be modeled using agent-based modeling, a paradigm well-known for modeling Complex Adaptive Systems (CAS). While agent-based models are also well-known to effectively model CAS, often times models can tend to be ambiguous and the use of purely text-based specifications (such as ODD) can make models difficult to be replicated. Previous work has shown how formal specification may be used in conjunction with agent-based modeling to develop models of various CAS. However, to the best of our knowledge, no such model has been developed in conjunction with AIDS. In this paper, we present a Formal Agent-Based Simulation modeling framework (FABS-AIDS) for an AIDS-based CAS. FABS-AIDS employs the use of a formal specification model in conjunction with an agent-based model to reduce ambiguity as well as improve clarity in the model definition. The proposed model demonstrates the effectiveness of using formal specification in conjunction with agent-based simulation for developing models of CAS in general and, social network-based agent-based models, in particular.

# *1. Introduction*

As noted in reports by the World Health Organization (WHO) (WHO, 2011), the Acquired Immunodeficiency Syndrome (AIDS) contributes *0.8%* (i.e., approximately *34* million people) to the global burden of diseases. While the mode of spread of AIDS and its prevention strategies are quite well-known, a clear indication of the complexity of HIV/AIDS spread can be noted by its continual prevalence in regions globally. The causative agent of AIDS is a retrovirus termed as the "Human Immunodeficiency Virus (HIV)". WHO reports note that areas such as Sub-Saharan Africa have nearly 1 in every *20* adults (*4.9%*) living with HIV thereby accounting for *69%* of the people living with HIV worldwide.

The variation in the emergent pattern of the AIDS/HIV pandemic in global populations is primarily due to the numerous and complex interactions of a large number of heterogeneous entities or agents[1] (Dean & Fenton, 2010) and populations. One of the pioneers of the theory of "Complex Adaptive Systems" (CAS), John Holland (JH Holland, 1992) forecasted that AIDS spread can be modeled as a CAS; systems which are traditionally associated with the observance of unusual patterns, termed as "emergent behaviors" ([Boccara, 2010](); [Muaz A Niazi & Amir Hussain, 2011]()). Considering AIDS as a CAS would imply that we can expect HIV/AIDS to exhibit emergent phenomena as a result of non-linear interactions of its constituent agents. However, to the best of our knowledge, although such patterns are expected and predicted by CAS researchers (such as Holland), such patterns have not been modeled or highlighted much in existing literature which has traditionally focused on studying AIDS primarily from the healthcare perspective.

---

[1] "agent" being used here as a technical term, which can refer to a person, a group, a cell or even a particular country or sub-population (M. Niazi & A. Hussain, 2011).

In this paper, the research question addressed is as to how a formal specification model can be coupled with an agent-based model of AIDS/HIV spread for examining HIV allowing for the discovery of emergent patterns. We use existing techniques such as studied in previous literature ([M. A. Niazi & A. Hussain, 2011a]()) to develop a Formal Agent-based Simulation Framework of HIV/AIDS spread (FABS-HIV). The structure of the rest of the paper is as follows: Section 2 presents background. In section 3, first a formal specification framework of the HIV/AIDS spread "FABS-HIV" is described followed by the description of an agent-based model developed in line with this formal specification. Section 4 presents implementation details of the agent-based model as well as simulation experiments, results and discussion. In the final section, the paper is concluded and future research directions are outlined.

## *2. Background*

This section first provides background and related work about modeling HIV/AIDS spread, formal specification and Agent-based.

### *2.1. HIV Disease Spread*

The HIV/AIDS epidemic has received considerable interest in the last two decades. With specific journals dedicated to different aspects of studying HIV/AIDS, there is considerable information available on both the actual virology of the epidemic as well as on infection spread. In addition, it is well-known how protection from HIV/AIDS can be achieved. However, in spite of this considerable amount of literature and wide-spread knowledge, the AIDS epidemic still poses a serious problem with worldwide prevalence.

Complex dynamics in AIDS were suggested by Jones and Handcock for disease spread associated within sexual social networks ([Jones & Handcock, 2003]()). A key problem in investigative studies of HIV/AIDS is in the collection of real-time data of disease spread due to challenges such as the taboo nature of the way it spreads in populations and subsequently how HIV transforms into AIDS and the hitherto unknown exact factors which causes the HIV virus to behave in a seemingly unpredictable manner (Chin & Kroesen, 1999).  Thus, while not all cases of HIV spread can be monitored easily, post-infection data can be considerably scarce (Allen et al., 2003).  This can result in developing results based on speculations and are limited by the extent of the sampled data. Thus, even the guidelines for determining non-probabilistic sample sizes are virtually nonexistent in most affected countries, mentioned by Guest et al. (Guest, Bunce, & Johnson, 2006). Whereas Pollack et al. (Pollack, Osmond, Paul, & Catania, 2005) note that risk behavior can be overestimated in the case it is defined broadly. In other words, while the existing schemes of data collection have given interesting results, there is actually a possibility of considerable room for further exploration and investigation in how different studies are practically structured and conducted.

## 2.2. Formal Specification

Formal specification is a type of formal methods employed for mathematical modeling of real-world systems  (Hall, 1990). While modeling a system, it is necessary to ensure that all the system components have been incorporated at the desired level of abstraction. Thus it is usually demanding and very effective strategy to use formal methods for the correct implementation of the system under investigation. The use of formal specification in the design phase of system modeling is suggested by Bowen  (Bowen, 1995) for earlier debugging of the system. "Z" is an ISO standard formal specification language developed in 1970s at Oxford University  (Woodcock & Davies, 1996). It has been used to model large scale real systems including IBM CICS (Houston & King, 1991) and cardiac pacemaker (Verdasca et al., 2005). Formal semantics of Z specification includes mathematical notations based on set like and first order predicate calculus. These notations are represented using formal structure known as "schemas" in Z specification.

In this paper, we have employed a Z-based formal specification model for HIV/AIDS modeling as part of FABS, an established modeling framework which has previously been shown to effectively model and analyze different types of CAS  such as sensing emergence such as flocking near Wireless Sensor Networks (M. A. Niazi & A. Hussain, 2011a, 2011b) and emergent behavior of citations of scholars as they progress in their research careers (Hussain & Niazi, 2013). Further to the justification of using formal specification is the fact that in spite of various modeling methods applied to investigate the HIV/AIDS spread dynamics there is lack of effective monitoring of the AIDS disease perhaps because of missing specification related to the modeling of CAS in context of HIV/AIDS spread. The reason behind this dilemma needs to be investigated through a modeling paradigm which takes into account a formal specification model of the system as offered by Z formal specification language.

### 2.3. Agent-based Modeling

Agent based modeling (ABM) is a simulation paradigm which has close ties with actual scientific experiments thus making it a good technique for evaluation different paradigms (Li, Brimicombe, & Li, 2008). What makes ABM unique for modeling complex adaptive systems is the way an ABM is typically designed. ABMs have been used in domains as diverse as Biological Sciences (Bailey, Lawrence, Shang, Katz, & Peirce, 2009; Siddiqa et al., 2009), disease spread models such as for Dengue (Jacintho, Batista, Ruas, Marietto, & Silva, 2010), Social Sciences (Dennard, Richardson, & Morçöl, 2008), Economics (Cartier, 2004) and Computer Sciences (Muaz Niazi & Hussain, 2010; M. A. Niazi & A. Hussain, 2011a). ABM has been termed as a revolution in the prestigious journal "Proceedings of the National Academy of Sciences" (Bankes, 2002). Holland also notes that CAS can be better studied by the use of agent-based models (John Holland, 2006). A recent review by Niazi (M. A. Niazi, 2013) presents literature about modeling CAS using agent-based and complex-network based models.

## 3. Model Development

This section includes description of informal model concepts needed for modeling FABS-HIV. It further includes description of formal specification model developed using Z and subsequently transformed working agent-based model.

### 3.1. Informal Description of Model Concepts

Modeling of HIV/AIDS spread involves dealing with heterogeneous population subsets with complex intra and inter communication patterns. The infection starts in a closed population sub-group[2] sharing common disease spreading behavior usually termed as high risk category. These high risk categories are further classified according to the mode of infection transmission used by them for example Injecting Drug Users (IDU) share syringes for intravenous drug intake and sex workers communicate through physical interaction as a mode of infection transmission etc. These sub-groups become a core source to transmit the infection to other population sub-groups which might be intermediate or secondary contractors of the infection. The intermediate individuals directly contract the disease from high risk individuals and in turn become a source to spread it to the population strata which is not directly exposed to any of the sources of HIV infection (thus termed secondary). Thus a complete picture of HIV/AIDS spread system demand the modeling of both intra and inter communication pattern involved in these population sub-groups instead of individually modeling each category as is the case with previous studies.

We in this study will limit our model to a single high risk category i.e., Female Sex Workers (FSWs) for simplicity's sake. The FSWs interact directly with their clients. Their clients in turn interact with their social partners belonging to low risk population sub-group[2] which are further divided into two categories. These two categories include both the socially committed and non-committed partners. The interaction between individuals belonging to any of these sub-groups might or might not be using preventive measures, another important factor in determining the shape of infection pandemic. This informal description enlists the key concepts to explain how infection actually travels in population. The same pattern is replicable for other high risk categories which is the reason for limiting our model to single high risk category for conduction of exploratory analysis of interaction pattern.

## *3.2.    The FABS-HIV Framework*

The FABS modeling framework involve two phases of model construction. First a formal model of HIV/AIDS spread using a mathematical framework originally based on a formal specification language Z is constructed. This formal specification model aims to validate all the informal needs of the system under investigation at the selected level of abstraction besides providing the unambiguous representation of concepts. Next an agent based model based on the formal specification framework is developed as a proof of concept. Subsequent simulation experiments performed by means of setting a 100% condom usage in populations (as control setting) were used for validation. Our results based on 95% confidence interval demonstrate the existence of complex

---

[2] "Sub-group" refers to a subset of people in the population which share commonalities in terms of disease spreading behavior

emergent interaction pattern observed as backflow of HIV infection in FSWs population based on coupling habits of their clients. Thus, in general, our experiments demonstrate why there exists a need for a realistic division of sub-populations in HIV studies in line with the local population sexual habits and behavior implying AIDS/HIV spread to be modeled as CAS. Our results suggest that instead of simply dividing populations between high risk and low-risk categories, studies should examine populations based on topological structure exhibited within and among the sub-populations. Following a more logic division in HIV studies could thus assist in a better evaluation and hopefully development of better mechanisms for controlling the spread of the HIV/AIDS spread worldwide.

As such the formal agent-based simulation framework consists of three types of parameters i.e., hard-coded, programmable and measured. The hard-coded parameters allows for the representation of key concepts based on the facts associated with the system, programmable parameters allow for state initializing of the system and measurable parameters allow for the subsequent monitoring of behavior dynamics.

The model is described using some key concepts required for modeling of FABS-HIV. The model dealt with a population consisting of various sub-populations that were categorized on the basis of risk behavior. These included one HIV/AIDS high risk population; i.e., Female Sex Workers (**FSWs**) in this case and three seemingly low risk populations: 1) heterosexual men (termed "**primaries**") being the people who have occasional interaction with the FSWs as their clients), 2) heterosexual women (termed "**secondaries**") since they do not have any direct interaction with the FSWs and acquire infection via primaries being their socially committed partner, and 3) heterosexual women who are not committed partners of the primaries, termed **exsecondaries**. The next key concept is to model the interactions between all these sub-population categories. The interactions required for modeling of FABS-HIV can be concluded as follows:

    a) The interaction between primary and secondary

b) The interaction between primary and exsecondary

c) The interaction between primary and FSWs

All these interactions are based on the moral behavior of primary population which can be described using two more concepts i.e., commitment level and adopting preventive measures such as condom usage. A high commitment level of a primary with its socially committed partner is indicative of keeping away from all other types of partnerships and in turn the infection itself and vice versa. Similarly, using the preventive measures like condom usage is also an indicator of partially staying away from the infection. We assume some basic knowledge of Z notation for simplification.

### 3.2.1. Sets

Sets have been used in the model to declare the user defined types in the model. We started by defining the given sets.

The set *"PERSON"* is the set of all persons in a given population:

[PERSON] == {x:ℕ | x>=0 }

Each person in the population carries different attributes regarding population sub-group category, infection status, gender, and their social status for coupling with their partners. Each of these concepts was adequately defined by associating different attributes with each *PERSON* entity. The free type definition *"PERSONTYPE"* depicts the different types of individuals in a given population; i.e., **fsw, primary, secondary and exsecondary.** A *PERSON* entity can belong to any one of these types. Formal definition of *PERSONTYPE* is as follows:

[PERSONTYPE] == {*fsw, primary, secondary, exsecondary*}

The free type *"PERSONSTATE"* was used to declare the health state of the individual which could either be *infected* or *uninfected:*

[PERSONSTATE]=={infected, unifected}

The next free type *"PERSONGENDER"* was defined as a means of declaring the gender of the individual which could either be *male* or *female*:

[PERSONGENDER]=={male,female}

The next set *"INFECTED"* defined all the infected individuals. It belongs to both programmable and measurable component of the model. It is used to monitor the system state encompassing infected population only, at a given time point:

[INFECTED]=={n:INFECTED|0<n<#persons∧ personstate=infected }

where *#persons* is a count of total no of persons in the population.

Likewise the set *"PARTNER"* is the set of all partners of a person. The *persontype* of *partner* was restricted to either *primary* or *secondary* since the *partner* is used to model the relationship of socially recognized commitment between these two entities:

[PARTNER] = = {x: PARTNER| persontype= primary ∨ persontype= secondary}

The set *"COUPLE"* was used to declare the pairs of sexually interacting individuals in the population. Each couple models the interaction of *primaries* with any of the other female sub-populations. Therefore, the predicate of the couple was restricted to at least one *primary*:

[COUPLE]== {x,y:COUPLE | x∈primary ∨y∈primary}

The set *"$\mathbb{F}$"* was defined as the set of natural numbers greater than or equal to 0 and less than or equal to 100. The natural numbers between 1-100 provided a scale for adjusting user-desired variables like commitment and condom usage as a property of the primary population. It was also a programmable parameter:

$[\mathbb{F}] == \{ x: \mathbb{N} \mid 0 < x \leq 100 \}$

Several user-defined global variables were declared with axiomatic declaration as programmable parameters. The variables *"maxprimary, maxsecondary, maxfsw, maxinfectedfsw, maxexsecondary, and tobecoupled"* were used for declaration of the number of *primaries*, *secondaries*, *FSWs*, the *infected FSWs,* the *exsecondaries, and* the number of pairs to be formed between *primaries* and *secondaries,* respectively. These global variables were provided for customized modeling of the initial population:

$$
\begin{array}{|l}
\textit{maxprimary}: \mathbb{N} \\
\textit{maxsecondary}: \mathbb{N} \\
\textit{maxfsw}: \mathbb{N} \\
\textit{maxinfectedfsw}: \mathbb{N} \\
\textit{maxessecondary}: \mathbb{N} \\
\textit{tobecoupled}: \mathbb{N}
\end{array}
$$

### 3.2.2. State Schemas

A state schema is a special structure used for the declaration of the system state. Each system state requires a specific set of objects, variables, entities and functions to ensure that all system requirements are being met. The states of FABS-HIV might be summed in three concise declarations:

a) State declaration for each population sub-group

b) State declaration of socially committed partnerships

c) State declaration for coupling action based on heterogeneous male population behavior

As a first step, the state schemas for all the sub-populations were declared to accommodate for their heterogeneity. The first population subtype object declared was *Fsw*. Every *Fsw* was declared with a set of attributes; i.e., *persontype, gender* and *state*. The predicate function restricts the person type to *Fsw*, and the gender to *female*:

```
┌─ Fsw ─────────────────────────────
│ type : PERSONTYPE
│ state : PERSONSTATE
│ gender : GENDER
├───────────────────────────────────
│ persontype = fsw
│ gender = female
└───────────────────────────────────
```

The next population subtype defined was *Primary.* The predicate function restricts the person type to *Primary*, gender to *male* and declares it with a socially committed partner which must belong to *Secondary* subtype:

```
┌─ Primary ─────────────────────────
│ type : PERSONTYPE
│ state : PERSONSTATE
│ gender : GENDER
│ apartner : PARTNER
├───────────────────────────────────
│ persontype = Primary
│ gender = male
│ apartner = Secondary
└───────────────────────────────────
```

The next sub-population declared was **Secondary**. The predicate function restricts the person type to **secondary**, gender to **female** and declares it with a partner who must belong to **primary** subtype:

```
┌─ Secondary ──────────────────────────────┐
│ type : PERSONTYPE                        │
│ state : PERSONSTATE                      │
│ gender : GENDER                          │
│ apartner : PARTNER                       │
├──────────────────────────────────────────┤
│ persontype = Secondary                   │
│ gender = female                          │
│ apartner = Primary                       │
└──────────────────────────────────────────┘
```

**Exsecondary** is an object of type **Secondary**. This category is used to depict the females who belong to low risk population with no socially committed partner at given time point. However they might be involved in occasional sexual encounters. The predicate function restricts the *exsecondary* to a null partner:

```
┌─ ExSecondary ────────────────────────────┐
│ exsecondary : Secondary                  │
├──────────────────────────────────────────┤
│ partner= Null                            │
└──────────────────────────────────────────┘
```

After defining all the population sub-groups adequately, we formulated the state schema **Partners** to store the pointers of the partners as a couple. This schema tends to create a pair of socially selected partners necessarily should be a *primary* (a male) and a *secondary* (a female). The aim of this object is to only declare the socially committed partners in a population. It has nothing to do with the coupling (sexual interaction) between these partners yet. The term coupling refers to the sexual act between heterosexual individuals.

The set $\mathbb{F}$*PARTNER* depicts a set of finite elements. Each element of this set consists of two entities, thus forming a pair. Each pair consists of two elements which holds pointer to each other. Each created pair can be called by using notation $\mathbb{F}$*PARTNER*. However, the elements of each pair are accessible using notation *PARTNER*. Three conditions were checked in the predicate. First condition checked for both elements of a pair to either belong with *Primary* or *Secondary*. Second condition monitored that no input entity is reusable i.e., an individual who has been used to form a pair is not used again in another partnership. Third condition implied that the pair formed uses the user-defined range for input values. The numbers *maxprimary* and *maxsecondary* depict the global variables used to declare the user defined values for the maximum number of *primaries* and *seconadries*, respectively. The total number of pairs to be formed was declared using a global variable *tobecoupled.*:

*Partners*
*x,y*: PERSON
*apair*: 𝔽PARTNER

($x \lor y$)∧($Primary \lor Secondary$)
$x \land y \notin PARTNERS$
$\#primary \leq maxprimary$
$\#primary \leq tobecoupled$
$\#seconadry \leq maxsecondary$
$\#secondary \leq tobecoupled$

The next state schema **Link** modeled the heterosexual sexual interaction between two individuals. Two variables are needed to model such an interaction. In the interaction, to model a male population one would always need a *primary* and to model a female population, a person could belong to any of the female sub-populations. Secondly the primary would always go for its own *secondary* partner from *secondary* sub-set. The purpose was achieved by calling three variables. Two of these were the elements of a pair created through *Makepartner* schema. It was depicted by notation *x:PARTNER and y:PARTNER.* Another *person* variable was used to model rest of two female subpopulations and restricting it to belong to either of two female populations i.e., *fsw* and *exsexondary* in the predicate. The variable *acouple* held the pointers to the coupling individuals in the pair. Two variables commitment and threshold-commitment declares the user-defined value of commitment level associated with primary population and a threshold value for implementation of

commitment preference rule:

```
┌─ Link ─────────────────────────────────
│ p1 : PARTNER
│ p2 : PARTNER
│ p3 : PERSON
│ acouple : 𝔽COUPLE
│ commitment, threshold-commitment : 𝔽
├────────────────────────────────────────
│ P3 ∧ ( fsw ∨ exsecondary)
│ commitment< threshold-commitment<=commitment
└────────────────────────────────────────
```

The next step is initialization of the declared states to describe the system state when it is first started. The initializing elucidates the working of concepts as hard-coded parameters i.e., population sub-groups representation and programmable parameters i.e., partnership declaration and coupling frequency etc.

Each *fsw* was a heterosexual female who was initially uninfected. The state schema *fsw* was initialized as *InitFsw* with *type* fsw, *state* uninfected, and *gender* female:

```
┌─ InitFsw ──────────────────────────────
│ Fsw
├────────────────────────────────────────
│ type = fsw
│ state = uninfected
│ gender = female
└────────────────────────────────────────
```

Each *primary* was a heterosexual male who was initially uninfected with no socially committed partner. The state schema *primary* was initialized as *InitPrimary* with *type* primary, *state* uninfected, *gender* male, and a null partner:

```
┌─ InitPrimary ─────────────────────────────
│ Primary
│──────────────────────
│ type = primary
│ state = uninfected
│ gender = male
│ apartner = NULL
└───────────────────────────────────────────
```

Each *secondary* was a heterosexual female who was initially uninfected with no socially committed partner. The state schema *secondary* was initialized as *InitSecondary* with *type* secondary, *state* uninfected, *gender* female and a null partner:

```
┌─ InitSecondary ───────────────────────────
│ Secondary
│──────────────────────
│ type = secondary
│ state = uninfected
│ gender = female
│ apartner = NULL
└───────────────────────────────────────────
```

Each *exsecondary* was the element of *secondary* population with no socially committed partner. The state schema ex*secondary* was initialized as *InitExsecondary* with *type* secondary, *state* uninfected, *gender* female and a null partner:

```
┌─ InitExsecondary ─────────────────────────
│ Exsecondary
│──────────────────────
│ type = secondary
│ state = uninfected
│ gender = female
│ apartner = NULL
└───────────────────────────────────────────
```

The *partner* schema was initialized as *InitPartner*. The partnership between the *primary* (declared as variable *x*) and *secondary* (declared as variable *y*) sub-populations as socially committed partners was initialized in this schema. The number of *primaries* and *secondaries* to be declared as socially committed partners was chosen through the user-defined value of global variables *maxprimary* and *maxsecondary*. The *maxprimary* and *maxsecondary* in this schema were set to *#maxprimary* and *#maxsecondary* respectively. There was no value stored in the couple yet; therefore, the value for *acouple* was set to an empty set($\varnothing$).

*InitPartner*
―――――――――――――――――――
*Partner*
―――――――――――――――――――
*x=Primary*
*y=Secondary*
*maxprimary = #maxprimary*
*maxsecondary =#maxsecondary*
*tobecoupled=#tobecoupled*
*acouple=$\varnothing$*

The *Link* schema was initialized as *InitLink* schema. Initially, each variable was set to null as no coupling was performed.

*InitLink*
―――――――――――――――――――
*Link*
―――――――――――――――――――
*p1 = Null*
*p2= Null*
*p3=Null*
*acouple =$\varnothing$*
*commitment=Null*
*threshold-commitment=Null*

Next all the global variables were declared using axiomatic declaration. The user-defined variables *maxprimary, maxsecondary, maxfsw* and *maxinfectedfsw* were declared to be positive integers. The variables *coupledprimary* and *coupledsecondary* were declared to be zero initially as these variables will be updated by their respective usage in further operations. The *threshold-condomusage* and *threshold-commitment* were declared to be any value between 0 and 100 inclusive.

> *maxprimary*:ℕ
> *maxsecondary*:ℕ
> *maxfsw*:ℕ
> *maxinfectedfsw*:ℕ
> *maxessecondary*:ℕ
> *tobecoupled*:ℕ
> ─────────────────
> *maxprimary* => 0
> *maxsecondary*=>0
> *maxfsw*=>0
> *maxinfectedfsw*=>0
> *maxexsecondary*=>0
> *tobecoupled*=0

### 3.2.3. Operation Schemas

An operation schema is the special structure in which the change in the states of the system is illustrated. It performs all the functions on all components using predicate specified conditions. The results of operation schemas produce output of measured parameters needed for system analysis and monitoring.

The first operation of FABS-HIV was to initialize the population. For this purpose, a schema *SetupInitialPopulation* was defined. The schema used several global variables to deploy user-desired inputs. The global variables used in this schema were: *maxprimary, maxsecondary, maxexsecondary, maxfsws* and *maxinfectedfsws* taking input for the maximum no of *primaries*, *secondaries*, *fsws*, and the infected *fsws* in the initial population:

```
┌─ SetupInitialPopulation ─────────────────────────
│ primaries: PRIMARY
│ secondaries : SECONDARIES
│ exsecondaries: SECONDARIES
│ fsws: FSWS
│ infectedfsws: FSWS
├──────────────────────────────────────────────────
│ primaries' = #maxprimaries
│ secondaries' = #maxsecondaries
│ exsecondaries' = #maxexsecondaries
│ fsws' = #maxfsws
│ infectedfsws' = #maxinfected fsws
└──────────────────────────────────────────────────
```

The aim of the next operation was to declare the primaries and secondaries as social partners of each other. The operation schema was declared as "*MakePartners*". The operation on the *Partner* schema is performed which has been included with a delta(Δ) sign to show the change in it. Two objects of type *Primary* and *Secondary* were taken as inputs. Both of these individuals were checked for the condition that the count of coupled individuals in the simulation was still required or not according to the initialized values. They were additionally checked for not being the elements of *PARTNER* set. The total number of pairs to be formed was checked using several global variables i.e., *maxprimary, maxsecondary* and *tobecoupled* through predicate conditions. The fulfillment of all the conditions would result in making the *primary* and *secondary* partners of each other:

```
┌─ MakePartners ───────────────────────────────────
│ ΔPartner
│ x? : Primary
│ y? : Secondary
├──────────────────────────────────────────────────
│ x=Primary
│ y=Secondary
│ x∉PARTNER
│ y∉PARTNER
│ #primary < #maxprimary
│ #primary<#tobecoupled
│ #secondary< #maxsecondary
│ #secondary<#tobecoupled
│ apair'=𝔽PARTNER ∪{x?, y?}
└──────────────────────────────────────────────────
```

The next operation schema *Coupling* performed operation of coupling on the *Link* schema. A *primary* will always go for coupling with his partner if the *commitment* is greater or equal to the threshold value. Otherwise, it will go for the coupling with *FSW*. This interaction was modeled as a binary relation defined as *acouple* between the selected objects through commitment preference rules:

*Coupling*
─────────────────────
$\Delta Link$
$p1?, p2?:PARTNER$
$p3?:Fsw$
$commitment?, threshold\text{-}commitment? : \mathbb{F}$
─────────────────────
$p3?=Fsw$
$commitment? >= threshold\text{-}commitment?$
$acouple' = (p1?, p2?) \cup acouple$
$commitment? < threshold\text{-}commitment?$
$acouple' = (p1?, p3?) \cup acouple$

The next schema *ApplyCondomUsage* performed an operation on selected coupling pair from *Coupling* schema as an input to apply the condom usage rule. This operation modeled the infection spread in a given population based on the frequency of preventive measure deployed by the *primarys*. If the *condomusage* was less than the threshold value, the infection would spread and vice versa. The predicate function would check the *state* of both the persons in the pair and infect the other if one was infected; otherwise, the *state* would remain the same:

$\underline{Applyingcondomusage}$
$\Delta Coupling$
$x?$: COUPLE
$y?$: COUPLE
$condomusage?, threshold\text{-}condomusage?: \mathbb{F}$

$condomusage? >= threshold\text{-}condomusage?$
$x'=x?$
$y'=y?$
$condomusage? < threshold\text{-}condomusage?$
$x? \notin INFECTED$
$y? \notin INFECTED$
$x' = \{x?\} \cup INFECTED$
$y' = \{y?\} \cup INFECTED$

### 3.3. Agent-based Model of FABS-HIV

This section formally describes the translated agent-based model developed as a proof-of-concept based on the formal specification presented for FABS-HIV in order to study AIDS/HIV as a CAS. The model was developed using NetLogo (Wilensky, 1999), a freely available agent-based modeling tool that has previously been used extensively to model complex phenomena (M. A. K. Niazi, 2011).

### 3.3.1. Agents

Based on the formal specification model, we modeled three sub-populations: a high risk category i.e., FSWs, their clients termed primaries, and the partners of the primaries termed secondaries (Fig. 1). To differentiate between them, a color code was used where yellow color indicated FSWs, blue indicated primaries, pink indicated secondaries and red indicated HIV/AIDS infected individuals.

### 3.3.2. Interactions

The model used "*links*" to represent two people engaged in a sexual relationship. These links created *couples*. There were essentially two kinds of couplings envisioned in this model:

1) an interaction between FSWs and primaries

2) an interaction between primaries and secondaries. These again were of two types:

- between primaries and their socially selected secondary partners (e.g. married partner)

- between primaries and their other secondary partners (which are perhaps not coupled with any other primary)

To perform verification of the model, each time there was a coupling, a link was shown on the screen between the two agents as a connecting line as (Fig 1).

### *3.3.3. Model Input Parameters (Agent Behavior)*

Several input parameters were provided as controls in the model to adjust the behavior of the agents. The number of sub-populations could be adjusted through the sliders on the graphical user interphase (GUI) of the model (Fig 2). The differentiation between secondaries of the two types could be adjusted with the help of another slider named "ex-sec", where ex-sec were the low risk heterosexual women other than the long-term selected partners of the primaries. The coupling between primaries and secondaries was controlled through a coupling variable based on per-month frequency of interactions (coupling) and the coupling between the primaries and FSWs was controlled through a variable adjusted according to the average client coupling recorded for the FSWs (Fig. 2). Two key confounding parameters; i.e. the commitment level of the primaries and the condom-usage among the primaries was used to study their impact on the spread of HIV/AIDS. The values of these parameters could also be adjusted according to the sliding scales provided in the model (Fig. 2). All the user-desired adjustable parameters were chosen on the basis of reported statistical parameters used for empirical data collection of HIV/AIDS spread(UNAIDS, 2005) (UNAIDS, 2010).

### *3.3.4. Model Output Parameters (Counters)*

All changes occurring in the model are time-dependant. Several output monitors including plots and counters (display screens with relevant numbers) were used for displaying the updated variable state of the agents at a given time. The estimates of the number of infected FSWs, their direct partners and secondaries were displayed as output counters. Furthermore, the model also provided estimates of the number of non-committed-secondaries (the number of secondaries used in the simulation which were not committed with any of the primaries), non-committed-infected-secondaries (the number of infected non-committed secondaries), and the total-infected population, a number that gave the overall picture of the infection spreading in the population. More importantly, the model also displayed the numbers of FSWs which were back-infected from infected primaries, FSW-back-infected. In other words, these infections resulted due to the interaction of infected primaries with non-infected FSWs, a value that demonstrates the effects of the existence of a complex hidden network for HIV back-flow. Finally, the number of back-infections of primaries from secondaries; i.e., primaries-back-infected was also displayed individually. The infection curve for all sub-populations was displayed through a line plot on ABM screen.

## 4. Results and Discussion

In this section, we first give an overview of the implementation details followed by model calibration. Finally, we present experiments and discussion of results.

### 4.1. Implementation details

The desired input parameters were incorporated using input scales provided on the model GUI. Next the setup button was used to initialize the model based on adjusted parameter settings. Time control provided to run the simulation allows performing it either indefinitely or for a specific time period. These controls were provided in the form of two "go" buttons, which allowed the control of the simulation runs in respective manners.

### 4.2. Model Calibration and Methods

Our model allows for calibration of certain parameters using the real data which include the number of sub-groups of agent population, the coupling frequency of primaries and FSWs, condom usage, infection rate, and time period. We performed verification of our model using two confounding parameters i.e., commitment level of primaries and variation in condom usage among the primaries.

- ***Variation in commitment level of the primaries:** we experimented with a 0% commitment level since it would mean that the primary agent had no regard for his partner regarding sexual practice in regular life, while a 100% commitment level meant that the agent was totally committed to the partner. The value of percent commitment level was adjusted from 0 to a commitment level of 100 percent in increments of 20.*

- ***Variation in condom-usage among the primaries:*** *We experimented with a 0% condom-usage level since it would mean that the infection would not propagate at all while a 100% condom-usage level meant that the infection would propagate exponentially. The value of percent commitment level was adjusted from 0 to a commitment level of 100 percent in increments of 20.*

Each point in all experiments was tested in 50 individual simulations. Every time tick represents one month as the smallest collective time unit. The simulation results were rendered in a graphical format for easy understanding. An error line function was used to get a clear picture of the minimum, maximum and average values of the data (to a 95% confidence interval) to observe various trends (Figs. 3 and 4).

To ensure realism in the simulations, the input data was based on real sources (UNAIDS, 2005) (UNAIDS, 2010) for Pakistani population except the coupling parameter which was analyzed by incrementing it through equally spaced variation intervals. The results obtained from the various experiments conducted below thus reflect the trends found in the Pakistani population.

## 4.3. Experiment no 1: Test of Variation in the Commitment Level

The first simulation experiment tested the effect of commitment level of the primary partners of FSWs on the spread of infection among the various agents being tested. The commitment level of the primary population was studied since we felt that their actions could seriously affect the form and shape that an HIV epidemic could take within a population.

Simulation results, while varying the commitment level among the primaries, revealed that the average infection among FSWs increased if the commitment of primaries with the secondaries went down (Fig. 3A). On the other hand, when the commitment level of the primaries increased with their partners, the infection among the primaries went down drastically by nearly 50% (Fig. 3B). The results in Fig. 3C show that the infection in FSWs could actually be significantly increased due to a lack of commitment level among the primaries with their partners, revealing that the infection was actually due to a back-flow from the primaries to the FSWs. In Fig. 3D, the same trend was observed for back-infected primaries, revealing the possibility that it was not the infected FSWs passing the disease to the primaries as they were quite few in number; rather, the primaries could actually be mostly "back-infected" from the secondaries since the infections went significantly down with an increase in the commitment level. The results from Fig. 3, panels A and C confirm the existence of back-flow emergence patterns involved in the HIV/AIDS spread which could not be calculated through traditional methods and thus prove FABS-HIV to be presentable as an adequate modeling framework for analyzing a CAS like HIV/AIDS spread.  Interestingly, the rate of infected secondaries were seen to be very low when primaries had no regard of commitment perhaps because of high coupling rate with FSWs instead of their partners which gradually increased with higher commitment levels(Fig. 3E).  The infection rate within the secondaries decreased gradually after 50 percent commitment level, as expected.  This could perhaps be owing to the lesser infection flow in the secondaries because of low infection rates among primaries due to higher commitment and lesser couplings with the FSWs (Fig 3E).  Fig. 3F reveals the overall trend of varying commitment level which shows up as normal distribution in the form of a bell-shaped curve.

### *4.4.       Experiment no 2: Test of Variation in Condom Usage*

Next, the variation in condom usage on infection rates was studied among the various actors using simulation modeling with the assumption that any interaction using condoms would not propagate infection.  Thus, condom usage here assumed proper usage without any tears or exchange of fluids. The infection rate obtained by varying commitment level was plotted against the different types of agents to reveal trends in each population as before.

It was observed that infections in FSWs could be significantly reduced by nearly 50% if the condom usage of primaries was increased incrementally (Fig. 4A).  With a 100% condom usage, the simulation experiment could not stop, verifying the program.  The same effect was observed on the back-flow of infection to the FSWs (Fig. 4C), suggesting the possibility of infection from back-flow as mentioned above.

Effect of increase in condom usage among the primaries revealed a similar effect on infection as in FSWs.  However, the rate of decrease in infection rates initially among the primaries was observed to be slower than that observed among the FSWs (Fig. 4B) and the same was observed with the back-flow of infection to the primaries from the secondaries (Fig. 4D)

Fig. 4E reveals the effect of variation of condom usage on the overall number of infected secondaries. As expected, with a 100% condom use, the infections went to zero, demonstrating the verification of the simulation program.  Finally, the effect of increase in condom use on total number of infections revealed a similar trend (Fig.4F).  In all these calculations, the number avg-client-pmonth (Average Transactional Clients per month for FSWs) was also taken from the self-reported data.

### *4.5.Correlation with Related Work*

While, to the best of our knowledge, there is no other work which presents a FABS in the area of HIV/AIDS, in this section we present a comparison of our presented work with related work.

A review of AIDS propogation modeling using agent-based model has previously been presented in ([Tirado-Ramos & Kelley, 2013](#)). One of the earliest agent-based model of HIV spread among injecting drug users (IDUs) was conducted by Atkins et al. in 1996 to simulate the spread of HIV from an index case to other IDUs using contaminated needles ([Atkinson](#)). Newer agent-based models consider network topology such as geographic isolation, social norms and also sexual taboos such as by Osgood et al. ([Osgood, Moavenzadeh, & Rhee, 2006](#)). Such studies have shown that depending upon the mode of transmission of the disease agents, the epidemic can take on different profiles. As such, a variety of intervention and policy decisions may have to be used to counter the effects of epidemic transmission such as those observed in Taiwan, Botswana, India, and South Africa ([Nagelkerke et al.](#)). Teweldemedhin et al. ([Teweldemedhin, Marwala, & Mueller, 2004](#)) use an agent-based modeling approach to model HIV spread in populations. Alam et al. ([Alam, Meyer, & Norling, 2006](#)) use an agent-based modeling approach to study the impact of HIV/AIDS in the context of socio-economic stressors. Mei et al. ([Mei, Sloot, Quax, Zhu, & Wang, 2010](#)) use agent-based networks to explain HIV epidemic in homosexual males in Amsterdam. Nagoski ([Nagoski, 2006](#)) presents an agent-based disease diffusion model to study heterogeneous sexual motivation. All these models are encompassing small set of agent/entities and the interactions as well, thus not employing the full potential of ABM to model a CAS. A previous paper on modeling HIV/AIDS as a CAS has been presented by Niazi et al. in ([MuazA Niazi, Siddiqa, & Fortino, 2013](#)).

Other previous studies based on formal methods include a fuzzy mathematical model developed by Morio et al. ([Morio et al.](#)) to analyze the heterosexual HIV transmission in Japan. A mathematical framework of concurrent partnerships on the transmission of HIV has been presented by Morris and. Kretzschmar in ([Morris & Kretzschmar, 1997](#)). Other mathematical models developed for the understanding of the HIV/AIDS epidemic and its transmission have been reviewed by Boily et al. in ([Boily, Lowndes, & Alary, 2002](#)). These models however lack a formal specification or modeling HIV/AIDS spread as a CAS.

In short, to the best of our knowledge, previous work either did not expand the use of agent-based modeling to the extent of modeling AIDS as a CAS or else if models were presented, they did not employ the use of formal specification. Likewise, previously presented models using formal methods or mathematical models have not connected these models to agent-based models. The current paper can be considered as extension of previous work on AIDS by means of developing a formal specification model in addition to expanding agent-based modeling to study emergent patterns in AIDS spread.

## 5. Conclusions & Future Work

In this paper, our goal was to present an agent-based model of AIDS spread, modeled as a CAS, coupled and correlated with a formal specification model. In particular, however, as an outcome of extensive simulation experiments, we have noted how a more natural sub-division of populations might lead to discovery of complex **emergent behaviors** in HIV epidemic spread. The results demonstrate how a more natural division of population preferably in the form of networks might allow for the detection of complex, emergent patterns in HIV spread.

We have demonstrated the use of a formal framework in conjunction with agent-based simulation model as an effective method in order to study complex diseases like HIV/AIDS. This formal framework allows conducting an investigative study for analyzing a complex adaptive environment like HIV/AIDS spread effectively. First a formal specification model of HIV/AIDS spread was developed using Z which allowed for a clear and unambiguous representation of the underlying complex social network. Next the formal specification was translated into an agent-based model as a proof of concept.

Extensive simulation experiments presented in a 95% confidence interval, demonstrated that in the case of FSWs, there can be back-infection patterns originating based on sexual habits and partnerships of their clients with a different section of population. We have also shown how the model can be verified and validated by evaluating the use of protection in physical relations as a protective measure from disease spread. Sensitivity analysis provided a proof of the utility of dividing populations into further sub-populations. The discovered emergent patterns are demonstrated since HIV spread is closely tied with the realities associated with the taboos and norms of the human society. The traditional modeling techniques could not predict the existence of complex hidden patterns and thus FABS-HIV proves to be presentable as an adequate modeling framework for analyzing a CAS like HIV/AIDS spread.

We demonstrate theoretically that infection emergence can change considerably depending upon the variations in the high risk behavior of certain populations. Our results also confirm the significantly positive role of both condom usage and commitment levels on stopping this possible backflow of infection in FSWs. Positive awareness of both these factors can play a significant role in better understanding of the disease diffusion rates specific to a population and also in planning effective intervention strategies.

Although the formal framework presented here has been described in the context of FSWs, it provides a complete description of the complex patterns involved in the flow of HIV/AIDS. In the future, further research can be conducted on the use of complex network methods to investigate trends in AIDS spread. In addition, FABS-based models can be developed for other sub-populations based on the behavior of high risk individuals such as injecting drug users (IDUs).

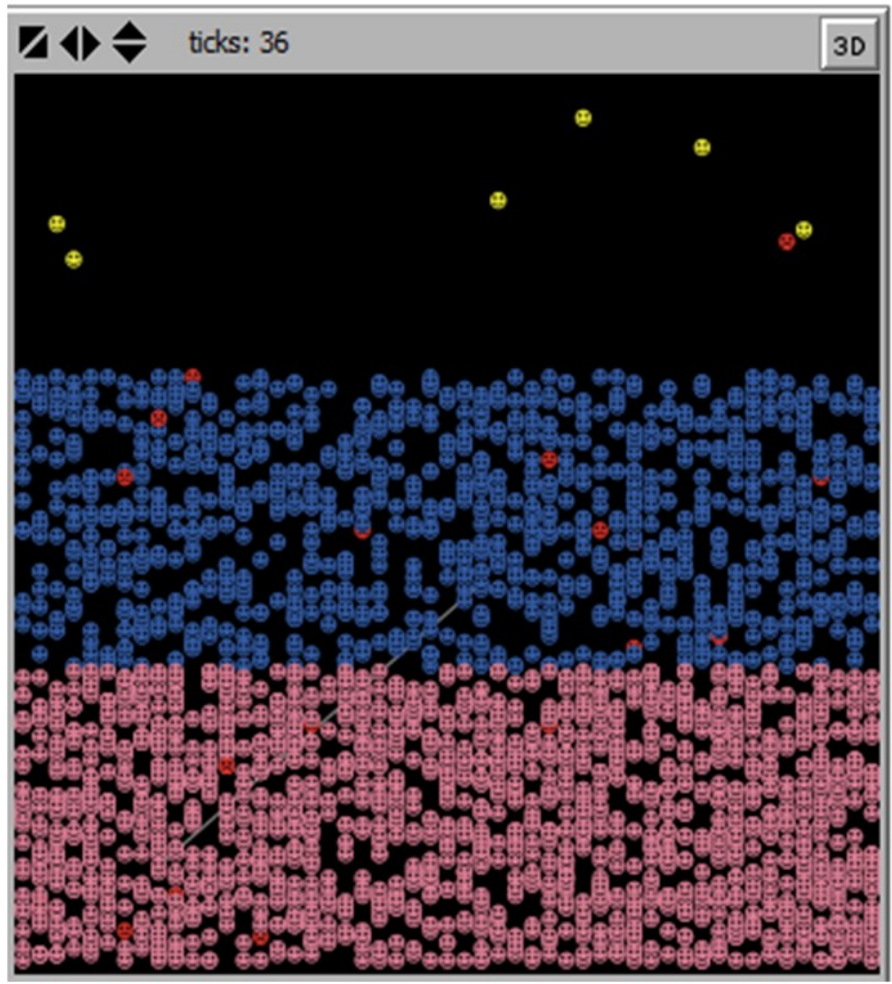

**Fig.1.** Screen shot of agents used during simulation modeling showing female sex workers, **FSWs** (yellow), infected **FSWs** (red), their clients, termed **primaries** (blue) and partners of primaries, termed **secondaries** (red). Sexual coupling between agents of different sub-populations during simulation is shown by vertical lines between agents. For each time tick (equally-spaced time interval), the interactions appear as lines between the relevant agents and then disappear in the next time tick for the sake of clarity.

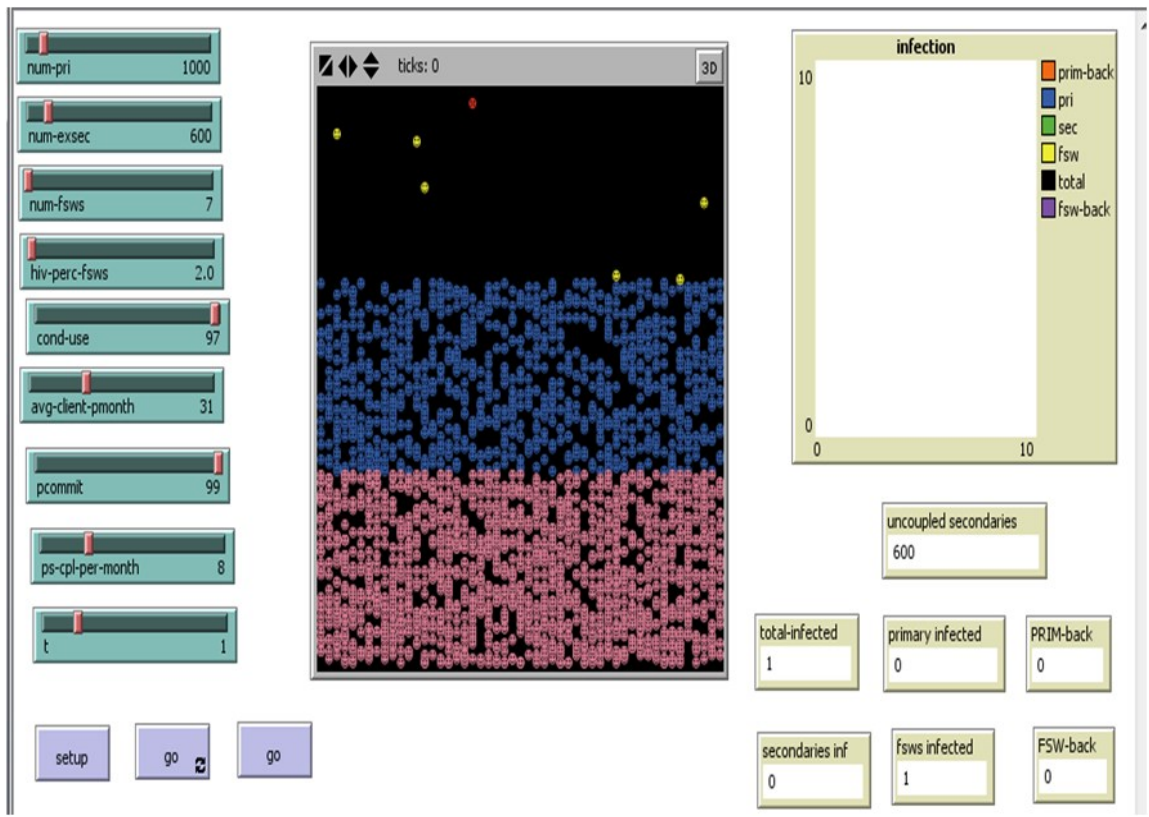

**Fig.2.** Screen shot of the graphical user interface (GUI) of the agent-based model of the FABS-HIV. The image shows sliders that allowed modulation of various sub-populations within the model as well as the level of confounding parameters like commitment levels and condom usage among the various agents.

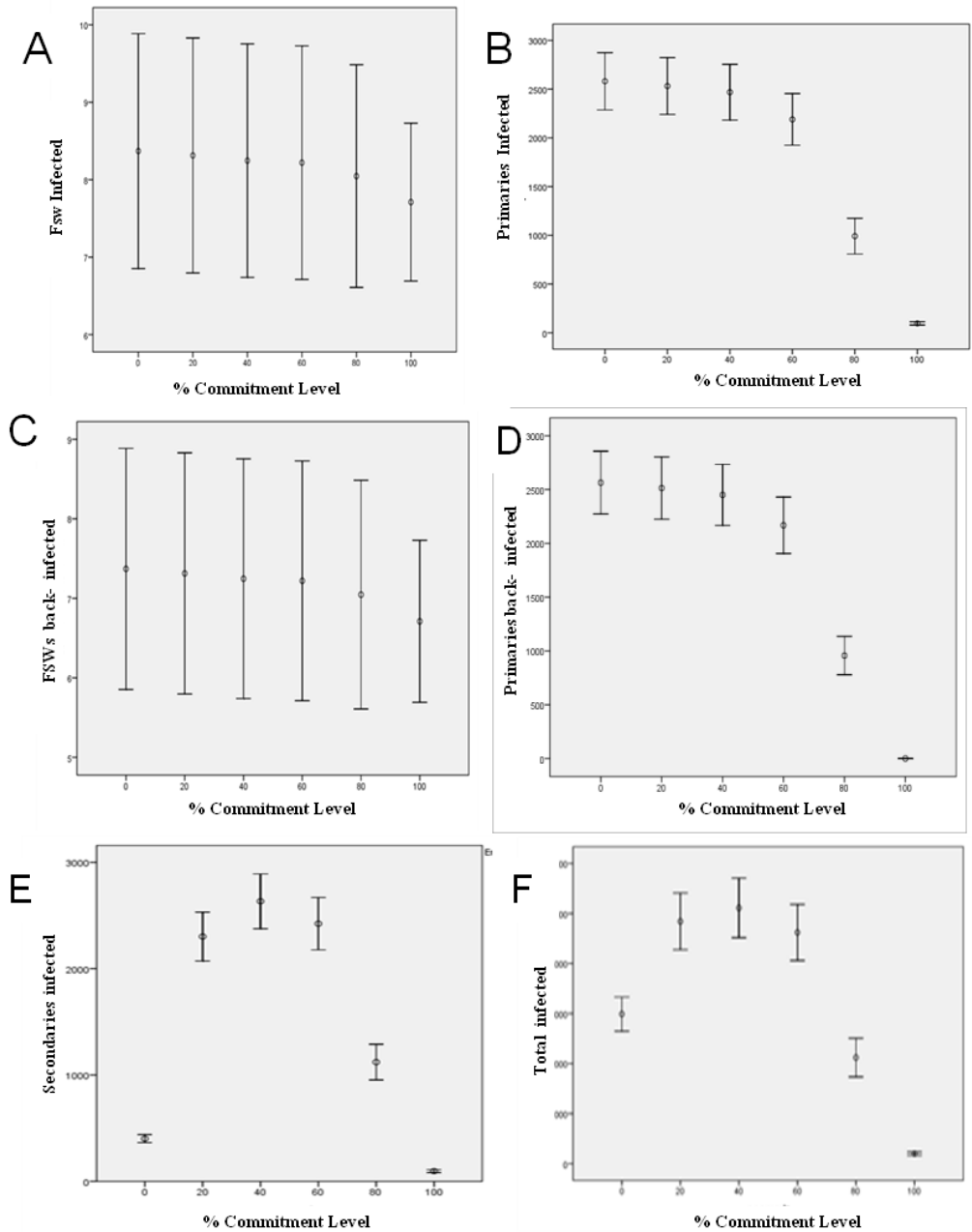

**Fig.3.** Results of simulation modeling by varying the commitment level among primaries. The graphs represent the estimated number of infected agents with increase in commitment levels: A) female sex workers (FSWs) infected, B) primaries infected, C) FSWs back-infected, D) primaries back-infected E) secondaries infected, F) total infected. The bars show the minimum, maximum, and average numbers estimated by simulation modeling using 50 rounds of simulations conducted per point.

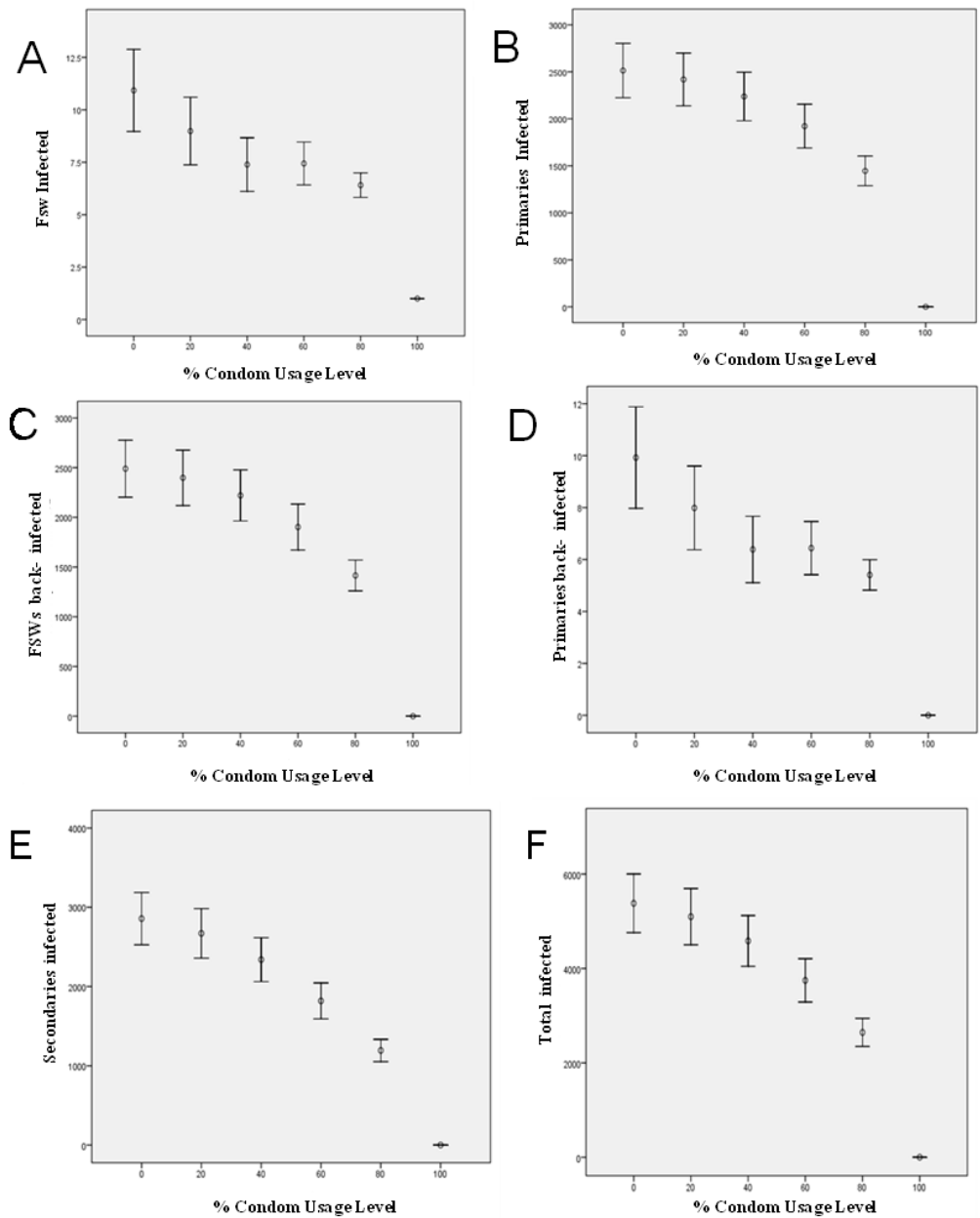

**Fig.4.** Results of simulation modeling by varying the condom use among primaries and secondaries. The graphs represent the estimated number of infected agents with increase in condom usage: A) female sex workers (FSWs) infected, B) primaries infected, C) FSWs back-infected, D) primaries back-infected, E) secondaries infected, and F) total infected. The bars show the minimum, maximum, and average numbers estimated by simulation modeling using 50 rounds of simulations conducted per point.